# *Catastrophic health expenditure and inequalities- a district level study of West Bengal*

*Pijush Kanti Das[1]*

1. Research Scholar, Indian Institute of Foreign Trade

**Abstract**

To achieve Universal Health Coverage (UHC), an important indicator under 'Sustainable Development Goals (SDGs)' requires inter alia that families who get needed health care do not suffer undue financial hardship as a result. In this study, we mainly aimed to estimate the incidence of catastrophic health expenditure (CHE) and analyze the extent of inequalities in out-of-pocket (OOP) health expenditure and it's decomposition according to gender, sector, religion and social groups of the households across Districts of West Bengal.

We analysed health spending in West Bengal, using National Sample Survey (NSS 71$^{st}$ round) pooled data suitably represented to estimate up to district level through multi stage stratified random sampling and interpenetrating sub samples. We measured CHE at different thresholds when OOP in health expenditure exceeded 10%, 20% or 40% of total household expenditure (WHO). Gini Coefficients and it's decomposition techniques were applied to assess the degree of inequality in OOP health expenditures and between different socio geographic factors across districts.

The incidence of CHE in West Bengal in 2014 with 10%, 20% and 40% threshold is in 30.1%, 22.0% and 15.1% households respectively. The incidence of catastrophic payments varies considerably across districts. Only 14.1% population of West Bengal was covered under health coverage in 2014. The inequality in OOP health expenditure for West Bengal has been observed with gini coefficient of 0.67. The same across districts varies between 0.453(Birbhum) to 0.752(Howrah). Prevalence of CHE and inequality was higher among poorer households. Gini decomposition of inequality in OOP health spending into 'between', 'within' and 'overlapping' part by subgroups of gender, sector, social group and religion group reveals varied results across districts, which may be used as policy prescriptions.

Based on the findings from this analysis, more attention is needed on effective financial protection for people of West Bengal to promote fairness, with special focus on the districts with higher inequality. Government has already initiated 'Swastha Sathi' Scheme to cover a large number of people, but which may be expanded to targeted socio-economic groups focusing on to diseases with high economic burden. The government may further increase in spending on services being provided at public health care facilities without compromising efficiency to reduce ever increasing reliance on private sector providers. This study only provides the extent of CHE and inequality across Districts of West Bengal but the causality may be taken in future scope of study.

## 1. Introduction

In recent years, the Universal Health Coverage (UHC) movement has gained global momentum, with the first-ever UN High-Level Meeting on UHC held in September 2019. The second UHC Forum was held in January, 2020 in Bangkok, aiming to enhance political momentum on UHC in international fora.

Health is also an essential part of the Sustainable Development Goals (SDGs). SDG 3.8 target aims to "*achieve universal health coverage, including financial risk protection, access to quality essential health care services, and access to safe, effective, quality, and affordable essential medicines and vaccines for all.*" It also reaffirms "health is a precondition for and an outcome and indicator of sustainable development" and recognizes UHC as fundamental to achieving the SDGs.

Universal health care means that all people have access to the health services they need without being exposed to financial hardship when doing so[1]. Poorer and vulnerable group of people within the population have the greatest need of protection from illness and disease[2]. Therefore, spending should be based on ability to pay and health services should be allocated according to need, ensuring that high out-of-pocket healthcare costs are reduced, and the associated impoverishing potential of poor health is reduced[1].

The National Health Policy 2017(NHP 2017)[3] in India envisages as its goal "the attainment of the highest possible level of health and well-being for all at all ages, through a preventive and promotive health care orientation in all developmental policies, and universal access to good quality health care services without anyone having to face financial hardship as a consequence. This would be achieved through increasing access, improving quality and lowering the cost of healthcare delivery".

But, Public expenditure on health in India remained stagnant near one percent of GDP till 2010, with an urban-centric policy orientation. India's public health expenditure was just 1.29% of GDP in 2019-20. In all recent years, the country lagged behind BRICs peers as well as developed nations. NHP 2017 targeted to achieve a health expenditure of 2.5% of GDP by 2025. While lack of effective access to health care by marginalised groups occurs in all societies, it is more pronounced in developing countries[4,5]. Moreover, in India, increasing private health care expenditure and out-of-pocket expenditure are making considerable financial burden on households[6]. The Social security in health expenditure is also remain low in India having on an average 67% of health expenditure is OOP in health expenditure in 2014, as per World Bank data.

West Bengal is one of the major State of India having characteristics of developing countries, experiencing rapid epidemiological transition, which involves dealing with both communicable and non-communicable diseases at the same time. Still, a huge number of West Bengal's population remain in severe poverty, which disproportionally affect women, SC, ST and minority communities due to unequal access to resources and employment opportunities and this continues to create disparities in health.

We, in this study focused on one measure of financial hardship that has been used widely in previous studies[2,7] typically referred to as catastrophic health expenditure(CHE). This measure is the official indicator for monitoring of UHC financial protection among the SDGs (indicator 3.8.2), with large expenditure suggested to be defined as 10% and 25% of total household expenditure. Although there is no consensus over the most adequate form to be used in studies regarding this subject, however, a low incidence of catastrophic payments could also mean people not getting (and not paying for) needed care.

On the other hand, Out-of-pocket (OOP) health expenditure is a suitable parameter for measuring inequality in health among different groups in societies[42]. OOP health Expenditure is households' expenditures for health-care service that insurers do not reimburse for them. The experimental results showed that OOP in health spending is significantly associated to CHEs[8].

Revision of literature identified few recent studies referring CHE and inequalities in health in India and West Bengal[9,10]. Although, no studies have analysed inequality in out of pocket expenditure for West Bengal for different socio-geographic factors and so that for district level. This paper addresses this research gap by examining the inter-district variation in households' health expenditure across different sub-groups.

This paper differs from earlier studies in a number of respects. First, we analyse the CHE and O.O.P. in health expenditure and related inequality using recent survey data NSS 71[st] Round(2014-15). Secondly, for analysing NSS data, probably this is the first time pooled data is being used, which have been designed through multi stratified random sampling in such a way that district level estimates can be obtained.

This paper has the following three broad objectives:

(1) To estimate the Catastrophic Health Expenditure (CHE) at district level as well as for whole State of West Bengal. Estimate of CHE has also been observed by Expenditure Class.

(2) Measurement of Inequality in Out-of-pocket expenditure in hospitalization cases considering the expenditure in Private care only, as public care in West Bengal are free.

(3) Decomposition of OOP health spending inequality by different subgroups of socio-geographic factors, viz.- gender, sector, social groups and religion

This knowledge is crucial for designing more effective health policies and programmes in West Bengal. For example, if among Socio-geographic factors inequalities in health expenditure due to caste found at a significant level, then programmes targeting caste will mitigate the inequality reducing burden in state coffer. District-wise assessment of CHE and inequality will help to give prioritize administrative and infrastructure reforms in particular districts first.

## 2. Methods
### 2.1 Data

The nationwide sample surveys on various socio-economic issues are being conducted on the basis of a set of "central sample" and "state sample" in each state. The state of West Bengal also participates in these surveys with equal matching sample basis. Due to inadequacy of sample size NSSO or State Directorate of Economics and Statistics(DES) does not release sub-state level estimates as such estimates are subject to high variability and less precision. One of the prime objectives behind collection of data by the DES is that the two data sets, one collected by NSSO and another collected by DES, may be pooled together to get more reliable estimates and benefit will be derived in the case of estimates at sub-state level like regions/districts. Therefore, pooling of central and state sample data may be considered as one way of tackling the problem of inadequate sample size for arriving at district level estimates.

This study presents estimates of various characteristics pertaining to CHE and OOP in health expenditure at State level in 2014 based on the pooled data of both central and state samples(collected and compiled by Socio Economic Survey Office(SESO),Bureau of Applied Economics and Statistics, Government of West Bengal) of NSS 71st round. The survey has been conducted during January-June, 2014, and total consists of 10,025 sample households. Sampling design consisted of 19 districts of West Bengal as reported in Table 2 in Annexure.

The entire data set is being finalized after satisfactory results in poolability test to nullify the agency bias. Since the parametric distribution of the sample mean is unknown, non-parametric tests such Run test and chi square tests were being adopted to test that the samples are coming from identical distribution function. Apart from these, parametric Z-Test were also done to test whether the expected value of estimates based on central sample data and that of estimates based on state sample data were equal or not.

**2.2 Estimating Catastrophic Expenditure Incidence:**

Catastrophic heath expenditure(CHE) occurs when a household's total OOP in health expenditure equal or exceed a certain threshold of his/her consumption expenditure. The threshold of 10% or 25% used in this study as suggested by SDGs. The variable on CHE is constructed as a dummy variable with value 1 indicating household with catastrophic expenditure, and 0 without catastrophic expenditure

$CHE_h = 1$, if $oop_h/aexp_h \geq 0.1$ or $0.2$ or $0.4$

$CHE_h = 0$, if $oop_h/aexp_h \leq 0.1$ or $0.2$ or $0.4$

Where, $oop_h$ is the annual out of pocket expenditure of h-th household and $aexp_h$ is the annual expenditure of that h-th household.

For, multistage stratified random sampling, calculation of appropriate weight/multiplier for a particular household is required for estimation purpose.

**Inequality and it's decomposition in Out-of-pocket health expenditure:**

For a long time, the analysis of economic inequality in a country was essentially a macro level exercise. The decomposition of overall income inequality by population subgroups and by income sources was introduced in the early 1980s[11,12]. The concept of decomposition of inequality signifies that population is partitioned into a number of sub-populations, the total inequality of the population can be expressed sum of the inequality within the sub-populations and of the inequality between them.

Two STATA files can be used: (1) INEQDECO (Jenkins, 2001[13]), estimating the full range of Generalized Entropy (GE) indices and providing decompositions for a subset of these indices by population subgroups, and (2) GINIDESC (Aliaga and Montoya, 1999[14]), decomposing Gini coefficients into between, within, and overlap parts, based on the algorithm by Pyatt (1976[15]). The decompositions is given by,

Gini(total) = Gini(within) + Gini(between) + Gini(overlap)

This paper critically assesses Gini decompositions by OOP in health spending and population subgroups following the GINIDESC techniques within the framework of Pyatt (1976).

In this expression, the between part accounts for the differences in mean incomes between the subgroups, and the within part depends on the inequality within each subgroup. The between part would be the only component if there was no variation in income/expenditure within the subgroups. The overlap part would be zero if there was no overlap between the income ranges of the various subgroups. The conventional between-group share is calculated by the ratio of observed between-group inequality to total inequality and interpreted to understand the inequality between two subgroups. The sign of this share has an interesting interpretation of higher income/expenditure for a particular sub-group.

The Gini coefficient and it's decomposition is easy to understand and also is the easiest way to report health-care financing inequalities to policy-makers.

## 3. *Results*
### 3.1 Catastrophic Health Expenditure

Figure, 1-3 at Annexure, presents District wise percntage of households having incidence to CHE. CHE has been categorized as three different levels of thresholds, eg 10%, 20% & 40% of usual annual consumer expenditure including both inpatient(hospitalization) and outpatient(OPD) health expenditure. The incidence of CHE in West Bengal at 10% is in 30.1% households. The same is 27.4% households in Urban area and 31.2% in Rural area. At the 20% threshold overall, Rural and Urban CHE households are 22%, 20.1% and 22.7% respectively. The incidence of catastrophic out-of-pocket payments varied strikingly across districts. At the 10% threshold, incidence ranged from 19.5% in Purulia to 41.4% in Paschim Medinipur. At the 20% threshold, the value for extreme districts are 12.2%(Purulia) and 26.1%(Bardhaman) respectively. Figure 4&5, represents a mapchart to showcase a variation of incidence of CHE across districts of West Bengal in 2014 at 10& 20% thresholds respectively. Malda, Bardhaman and Paschim Medinipur are most affected districts in both cases.

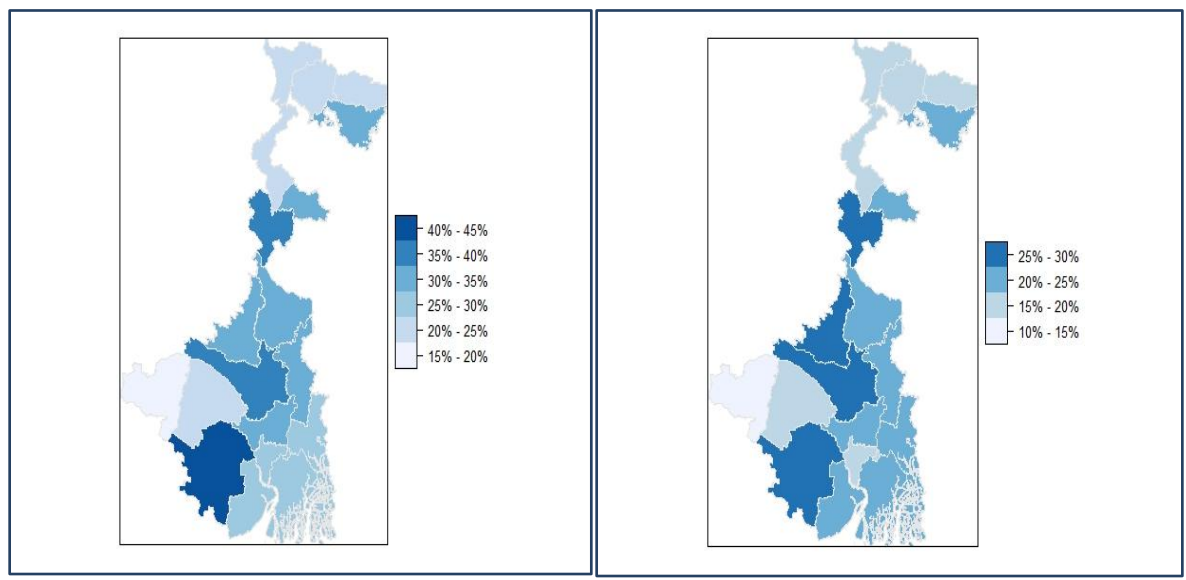

Fig 4: distribution of CHE at 10% incidence    Fig5: distribution of CHE at 20% incidence

Mean incidence of CHE at 10% threshold across districts was 29.2%(SD 5.5) and median was 29.1%(IQR 25.1-31.9). Incidence was inevitably lower at the 20% threshold (figure 2 at Annexure), with mean and median incidences of 21.4% (SD 4·1) and 21.6% (IQR 18.4-24.4). For calculating, extreme CHE with a threshold of 40%, still an incidence in 15.1% households, with a distribution of 7.4%(Purulia) to 18.6%(Bardhaman) has been observed. It can be easily observed increasing of CHE threshold did not improved in incidence in CHE substantially!

The rank correlation between the two catastrophic payment measures(10% and 20%) was 0·975, so for almost all part of low incidence at the 10% threshold compared with other districts (which could be interpreted as good performance) was mirrored by low incidence at the 20% threshold compared with other districts.

Table 1, shows the income-related distribution of the CHE incidence at different thresholds in 2014. The lower the household expenditure (proxy of income), the higher the incidence of CHE in all thresholds (10%, 20% and 40%). For example, during 2014 in West Bengal, when the threshold of CHE was 10%, the average incidence rates were 31.9%, 30.7%, 27.2%, 25.8% and 24.7% in quintiles I (the poorest quintile), II, III, IV and V (the richest quintile), respectively.

| Table 1: % of Households according to MPCE Groups (poor to rich) and incidence of CHE at different thresholds (10%,20%,25%&40%) | | | |
|---|---|---|---|
| MPCE Code | 10% threshold | 20% threshold | 40% threshold |
| 1 (Poorest) | 31.94 | 23.98 | 16.61 |
| 2 | 30.72 | 21.45 | 12.47 |
| 3 | 27.25 | 19.71 | 13.74 |
| 4 | 25.79 | 19.64 | 13.92 |
| 5 (Richest) | 24.72 | 17.50 | 13.27 |

Source: Author's calculation

Health Scheme Coverage: Figure 6 depicts, only 14.1% people are estimated to be under health coverage in West Bengal in 2014. Significantly high health coverage at 99% confidence interval across value of districts has been observed at Birbhum(36.5%), Kolkata(26.2%), Hooghly(24.8%), Paschim Medinipur(24.7%) and Purba Medinipur(19.7%). Significantly low coverage has been observed in Coochbehar(0.7%), Uttar Dinajpur (3.9%), Jalpaiguri(5.2%), Dakshin Dinajpur (5.7%), Bankura(6.7%), Purulia(8.3%) and Darjeeling(8.8%), mostly northern part and western part and less developed districts of West Bengal

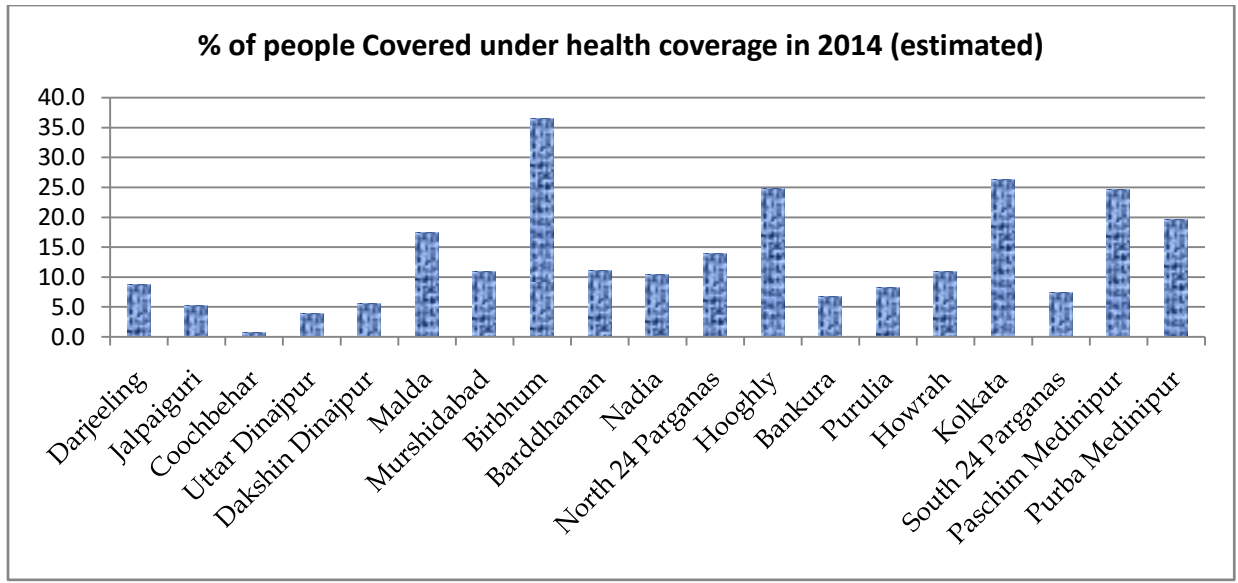

Figure 6: Percentage of people covered under health coverage (Author's calculation)

The spearman's rank correlation test between health coverage and incidence of CHE at 10% threshold, considering the same direction of events, provides a rank correlation of 0.413 with p-value 0.079, showcasing a significant correlation at 90% confidence.

## 3.2 Inequality in OOP health expenditure and it's decomposition across sub-groups:

Table 2 at Annexure, is important in two ways, one in calculation of overall inequality in OOP in health spending for all districts of West Bengal as well as overall for the State. Secondly, the interpretation of 'between inequality' measures to understand the inequality between different socio geographic factors. Though as we already discussed 'between inequality' must not be observed in absolute terms, rather it is appropriate to observe how much the between inequality explains among the total inequality across districts of West Bengal.

The result shows, the overall inequality in OOP in health expenditure in West Bengal is 0.67. The inequality in Rural area(0.64) is less than inequality in Urban area(0.71)(author's calculation not reported). The most affected districts are Howrah(0.752), Kolkata (0.743) and Darjeeling (0.726), on the other hand the districts with least inequality are Birbhum(0.453), Bankura(0.522) and Maldah (0.543).

The decomposition of inequality was mainly done on 4(four) factors, viz.- Gender(Male vs female), Sector(Rural vs Urban), Social group(SC&ST combined vs Others) and religion (Muslim vs Others). The share of Between inequality by these factors on overall inequality for the State as a whole are 7.9%, 15.5%, 8.8% and 5.7% respectively. But as negative sign is only to understand the direction, the absolute share may be used to measure the depth of inequality. The average of absolute percentage share of 'between' inequality components in total inequality considering 19 districts for gender, sector, social group and religion are 17.2%(CI: 11.6-22.8), 15.4%(CI:10.1-20.7), 14.6%(10.2-19.1) and 11.6%(CI: 5.7-17.5) respectively. The results shows on an average highest in between inequality share explained by male vs female.

The explanation of results by importance of factors are provided below. Significance in this part is at 99% confidence interval.

Gender: While decomposed on gender difference, observing in between inequality share in total inequality for respective districts and C.I. shows significant high male vs female inequality in Darjeeling, Purulia and Kolkata. Significantly low inequality has been observed in Bardhaman, Hooghly, Howrah and South 24 Pgns district.

Sector: After decomposition of inequality in OOP in health spending over geographic factors, the districts Darjeeling, Bardhaman, North 24 Pgns, Howrah and South 24 Pgns district have significantly high inequality between Rural vs Urban area, among which only Darjeeling have higher OOP in health spending in Rural area. As Kolkata consists only urban area, no in between part has been found. Districts with Significantly low sectoral inequality are Purulia, Uttar Dinajpur, Birbhum, Bankura and Purba Medinipur.  Though the problem of this interpretation is the location has been earmarked by place of residence and not by place of care, but some associated costs due to residential location of a person are inevitable.

Social groups: While decomposed in social groups, we combined SC and ST population together to make a group to observe inequality vs others. Some of the districts with significantly high inequality by measuring in between inequality share in total inequality are Darjeeling, Uttar Dinajpur, Birbhum, Nadia, Bardhaman and Paschim Medinipur. The significantly least affected districts in terms inequality between SC,ST population with others are Jalpaiguri, Dakshin Dinajpur, Maldah, Murshidabad, North 24 Pgns and Kolkata.

Religion: Decomposition over religion groups (Muslim vs others) reveals, significantly high in between inequality in Uttar Dinajpur, Nadia and Paschim

Medinipur District. Significantly low inequality in between Muslim and others have been observed in Darjeeling, Jalpaiguri, Purulia, Kolkata and South 24 Pgns district.

To understand the inequality, we also segregated the data with the persons with chronic disease and without chronic disease to account the effect of chronic diseases on OOP health expenditure. A two tailed T-test revealed no difference in mean inequality between two groups- with chronic disease and without chronic disease (t-stat=-0.746, p-value=0.46>0.05). On the other hand, t-test on mean inequality between two groups- overall and persons only with chronic disease reveals significant difference at 90% confidence (t-stat=1.76, p-value=0.089). For the population with chronic disease, inequality for the most 3 affected districts remains unaltered with gini coefficient 0.725, 0.704 and 0.698 respectively. The least affected districts are Purulia(0.356), Birbhum (0.369) and Uttar Dinajpur(0.424).

At disaggregated level of OOP health expenditure by different expenditure components reveals for hospitalization(in-patient) cases(Table 3 at Annexure), apart from package component, most of the expenditure goes into Package components(38.2%, CI:22.7-44.9 at 99%), Medicines(20.1%, CI:16.8-26.9) and Doctor's/Surgeon fees(10.2%, CI: 5.6-14.1) for the state as a whole. Package component in hospitalization cases are significantly burden in Hooghly, Darjeeling, Jalpaiguri, Kolkata and Howrah districts. Share of expenditure in purchasing medicines in OOP health expenditure is significantly higher in Uttar Dinajpur (40.8%), Dakshin Dinajpur(36.4%), Barddhaman(28.5%) and Birbhum(28.1%) with 99% confidence. Share of Doctor's/Surgeon' fees out of OOP health spending highest in Murshidabad(30.9%) followed by Coochbehar (20%) and which is statistically significant.

For outpatient cases(Table 4 at Annexure), expenditure in medicines (61.6%, CI: 51.6-66.9) followed by Doctor's fee(14.2%, CI:11.6-16.8) and diagnostic tests(10.3%, CI: 5-15.6) have higher burdens in OOP in health spending. People of Darjeeling, Uttar Dinajpur, Kolkata, Paschim Medinipur and South 24 pgns districts bear significantly high medicine expenditure in OP cases with 99% confidence. Share of expenditure in Doctor's fee for out-patient cases are significantly higher in Howrah, Purulia, Nadia, Purba Medinipur, Kolkata and Birbhum district.

Indirect cost associated with health expenditure like Transportation cost sometimes become a burden according to many Literature. In our case 6.3%( and 6.9% expenditure share is contributed by this cost for inpatient and outpatient case respectively for the State as a whole. But, a substantial variation of burden of transportation is seen across districts. For inpatient cases across districts, this expenditure share has been observed in a range between 1.2%-11.2%. The same is 2.3%-11.9% for out-patient cases. Purulia bears highest share and Kolkata share lowest share in both cases.

## 3.3 Discussion

To our knowledge, this is the first study using NSS pooled data to obtain district level estimates of CHE and decomposing health-related inequalities across different socio-geographic factors in India.

High incidence of CHE across districts has been observed in West Bengal in 2014, which required further analysis of OOP in health expenditure and related inequality. Looking at the various socio geographic factor, reveals a gender biasness towards male at the top of the ladder, followed by sectoral inequality disfavouring rural area, Caste biasness and least affected by religion inequality in the State of West Bengal in regards to health. The results of high CHE, may also be due to lesser number of people under health coverage by social security agencies or Government, high dependence on private providers due to lack of quality and facility of services under public health care along with income related burden due to unemployment or many other factors.

Though many studies observed higher health expenditure for women due to various reasons, like women have poorer health than men due to biological factors and more sensitive to health hazards as care giver to their family[16,17]. Our study revealed higher average expenditure for men across districts, with substantial inequality with women. It may happen still a large number of women are not being able to report their health hazards and lack in accessing better health services than their counterparts. Interventions that remove discrimination against women in health care, education, employment and income generating activities are crucial. The reduction of gender inequalities in these domains is also expected to improve women and children's well-being, owing to women's important role in food preparation and childcare.

A proxy of lower access to public health care over private care may be thought of in terms of percentage delivery. Figure 7 calculated from our data showed still a lack of use in public hospital across districts. Also, in an earlier study on Kerala, in Kerala which found that, even in government hospitals, households spent significant amounts of money on buying services outside the hospital[18].

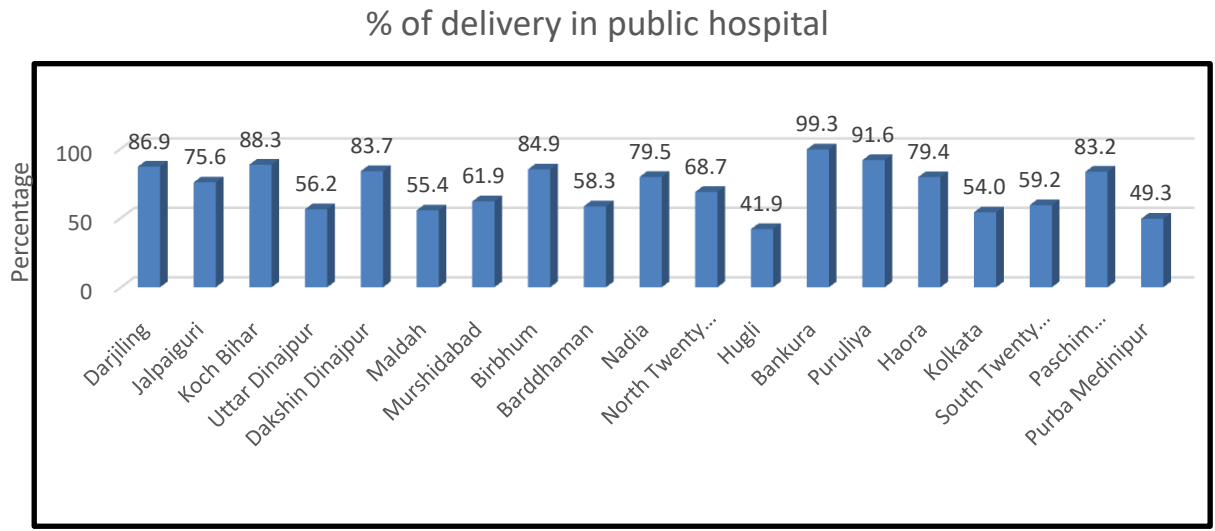

Figure 7: Percentage of delivery in Public hospital across districts (author's calculation)

In addition to this, inefficiency in service delivery in public health systems may be another contributory factor. For example, long waiting, neutral attitudes of health care workers and lack of essential medicines.

Other empirical findings suggest that high health expenditure for a household is not usually the result of one single disastrous event such as hospitalisation, but rather a series of events[19,20]. Our data reveals that in West Bengal, per year, a household's average expenditure on hospitalisation is only 34% of what it spent on OPD visits. It has been found in other parts of India that low expenditure on hospitalisation is due to low utilisation of hospital care, which may happen more for vulnerable groups. The concentration of hospitals mostly in urban areas and district headquarters is a barrier to access for rural populations, resulting ultimately in their low utilization of hospital care[20].

Financial risk protection is a core objective of universal health coverage[21]. Health coverage in West Bengal in 2014 found low across districts (figure 6). If a health care system is to meet the objectives of universal health coverage, total contributions should be based on ability to pay, and health care services should be allocated according to need, which means poorer people should receive greater health care benefits due to greater health care needs[21]. The financing capacity of the public sector must be improved, resources must be used efficiently, and an adequate system of the supply of consumables must be structured[22].

Disaggregation of health expenditure reveals substantial burden of different components specially Medicine fees, doctor's fees and diagnostic test on people varying across districts(Table 3 and 4 at Annexure). Appropriate administrative steps to reduce this variation may reduce inequality in health expenditure overall.

Mamata Banerjee led Government in West Bengal has ushered in sweeping reforms in healthcare system to achieve the goal of 'Universal health care' in the State since 2011-12 with several new schemes. Health Infrastructure has been increased by establishing Super Speciality Hospitals at the Districts. All treatments at Public Hospital has made free for all people. *Swasthya Sathi*, which was launched in 2016, has completed its three years of successful rollout. 1.5 crore families consisting of 7.5 crore population are covered under the scheme. The entire cost is borne by the State Government and there is no contribution required from the beneficiary who can avail coverage upto Rs. 5 lakh from large network hospitals of the State. The Swasthya Sathi Smart Card is issued in the name of woman member of the family. 1518 network hospitals and nursing homes are providing services under the scheme. Health Care benefit amounting Rs. 906.53 crore was availed by 8.99 lakh patients till 31.12.19. 'Zero Payment Drug Dispensing Slip' has been introduced in 2019-20, wherein patients get a slip mentioning the name of the medicines and the corresponding value of the medicines with 100% discount during the dispensing of the drug from the pharmacy counter at the Govt. Hospitals and West Bengal Clinical Establishments (Registration, Regulation and Transparency) Bill, 2017 was passed to control the increasing expenditure at private health care facilities[23].

Similar survey has been conducted NSS 75th Round(July,2017-June,2018), for which pooled data yet not been available. Comparison between NSS 71st round and 75th round is essential, not only to analyse the trend of CHE, rather to understand the effect of 'Swastha Sathi' scheme in terms of covered population under UHC and it's impact on CHE and inequality in OOP expenditure in health. The study may also be extended whether coverage under health plan is sufficient to the households or not.

*4. Conclusion*

In conclusion, our data (NSS 71st round pooled) estimated substantial incidence of CHE in West bengal across districts at all three households 10%,20% and 40%. The figures for West Bengal at respective thresholds by percentage of households are 30.1%, 22% and 15.1%. The incidence of catastrophic payments varies considerably across districts. We found intermediate rank correlation between percentage of people under health coverage with incidence of CHE at 10% threshold across districts(r=0.413, p=0.079). A low health coverage across districts has been observed as well as for the whole State. Only 14.1% population of West Bengal was covered under health coverage in 2014. In addition, an expressive inequality in OOP health spending has been observed across districts. The inequality in OOP health spending for West Bengal has been observed with gini coefficient of 0.67. The same across districts varies between 0.453(Birbhum) to 0.752(Howrah). Prevalence of CHE and inequality was higher among poorer households. Gini decomposition of inequality in OOP health spending into 'between', 'within' and 'overlapping' part by subgroups of gender, sector, social group and religion group reveals varies results across districts, which may be used as policy prescriptions.

Based on the findings from this analysis, more attention is needed on effective financial protection for people of West Bengal to promote fairness, with special focus on the districts with higher inequality identified by different factors. Government of West Bengal has already initiated 'Swastha Sathi' Scheme to cover a large number of people, but which may be expanded to targeted socio-economic groups focusing on to diseases with high economic burden. The government may further increase in spending on services being provided at public health care facilities without compromising efficiency to reduce ever increasing reliance on private sector providers. Attention must be paid to proper implementation and distribution of benefits from government expenditure. This study only provides the extent of inequality across Districts of West Bengal but the identification of determinants of CHE may be taken in future scope of study.

Though we think, this is the first use of pooled NSS data to provide such extensive analysis on CHE and OOP health spending across districts of a State in India, our study has some limitations.

Firstly, several problems occur in using these standard decomposition techniques. Only categorical variables can be used as grouping variables. It is not possible to incorporate covariates in the analysis. Also, if we have more than one grouping variable, these variables must be combined into one 'super' grouping variable. This will lead to an unmanageable number of categories.

Secondly, there are the limitations of NSS consumer expenditure data (which, as already noted, relate to private consumption) collected by the interview method and involving considerable amount of imputation work[24]. Therefore, we did not attempt to conduct inference around our state or district level estimates. Uncertainty around our estimates comes from both sampling and non-sampling error.

Third, the use of self-reported healthcare expenditure is a key weakness of a survey like NSS, which is also common to all previous surveys that used individual-level data to assess of socio economic conditions in India. It could be questioned whether individuals are able to accurately recall the amount they have spent on healthcare, which may have influenced the accuracy of the results.

Lastly, our analysis shows merely one dimension of UHC. A low incidence of catastrophic spending in some districts might simply reflect a situation in which only a few people get the health care they need because facilities are few or inadequate. data for both sides of the UHC coin need to be examined simultaneously.

Despite these limitations, this study is important in that it gives an understanding and quantification of the drivers and magnitude of CHE and inequalities in health expenditure across districts. This study well help to identify the districts for which population are disproportionately affected by CHE as well as health expenditure inequality for different sub-groups. Our results are by their nature associations and do not necessarily reflect causation. The study is more apt to take district level policy decisions by appropriate stakeholders.

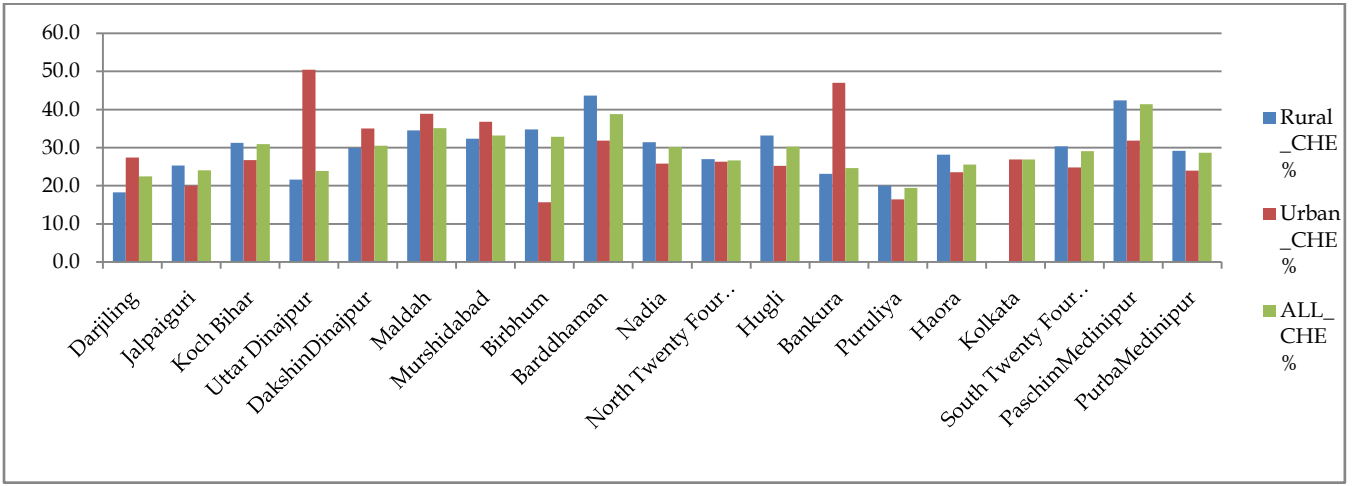

Figure 1: : Incidence of CHE by % of households at 10% threshold (Author's calculation)

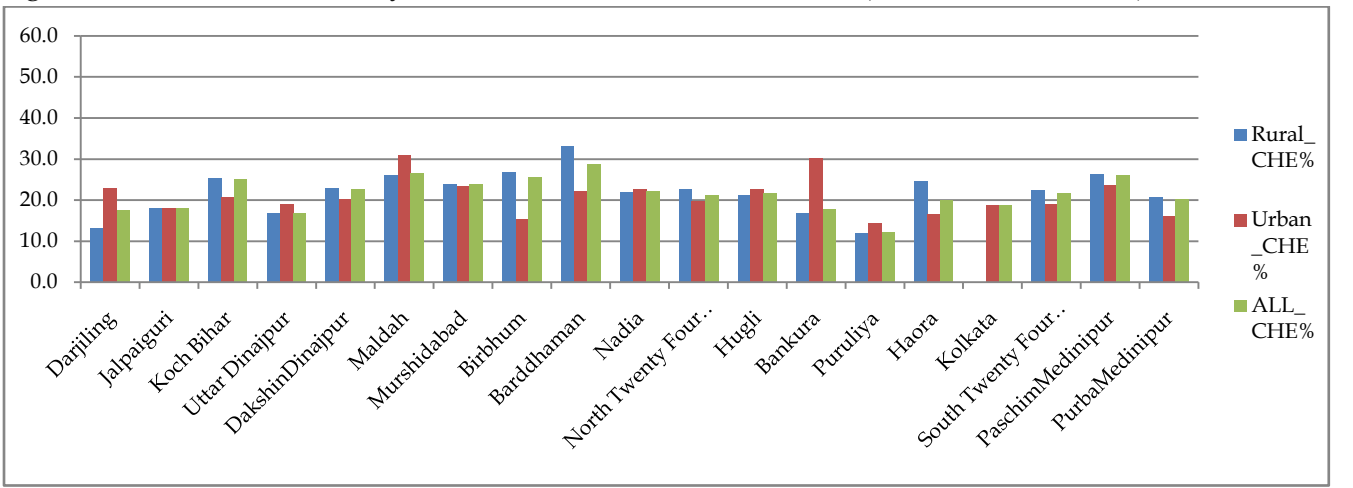

Figure 2: Incidence of CHE by % of households at 20% threshold (Author's calculation)

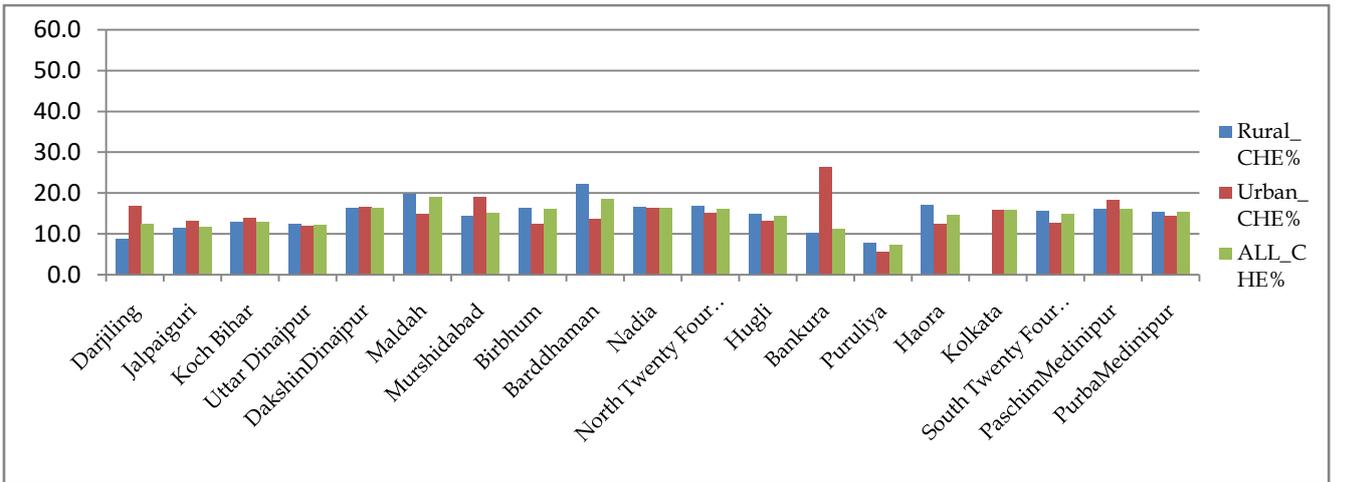

Figure 3: Incidence of CHE by % of households at 40% threshold (Author's calculation)

## Table 2: Gini inequality and it's decomposition across different sub-gropus

| State/Districts | Total Gini | Sex | | Sector | | SC,ST combined vs Others | | Religion: Muslim Vs Others | |
|---|---|---|---|---|---|---|---|---|---|
| | | Between | Within | Between | Within | Between | Within | Between | Within |
| Darjeeling | 0.726 | 0.309*** (42.6) | 0.365 (50.3) | -0.174*** (-24.0) | 0.453 (62.4) | 0.155*** (21.3) | 0.351 (48.3) | 0.004*** (0.6) | 0.722 (99.4) |
| Jalpaiguri | 0.565 | -0.109 (-19.3) | 0.276 (48.8) | 0.1 (17.7) | 0.307 (54.3) | -0.014*** (-2.5) | 0.285 (50.4) | 0.006*** (1.1) | 0.559 (98.9) |
| Coochbehar | 0.618 | -0.067** (-10.8) | 0.341 (55.2) | 0.111 (18.0) | 0.503 (81.4) | 0.11 (17.8) | 0.301 (48.7) | 0.04 (6.5) | 0.305 (49.4) |
| Uttar Dinajpur | 0.545 | 0.116 (21.3) | 0.28 (51.4) | 0.028*** (5.1) | 0.445 (81.7) | 0.156*** (28.6) | 0.255 (46.8) | 0.201*** (36.9) | 0.249 (45.7) |
| Dakshin Dinajpur | 0.601 | 0.11 (18.3) | 0.298 (49.6) | 0.081 (13.5) | 0.5 (83.2) | 0.022*** (3.7) | 0.428 (71.2) | 0.095 (15.8) | 0.284 (47.3) |
| Maldah | 0.543 | 0.096 (17.7) | 0.274 (50.5) | 0.092 (16.9) | 0.428 (78.8) | -0.009*** (-1.7) | 0.302 (55.6) | 0.067 (12.3) | 0.266 (49.0) |
| Murshidabad | 0.591 | 0.076 (12.9) | 0.313 (53.0) | 0.059** (10.0) | 0.372 (62.9) | -0.036*** (-6.1) | 0.377 (63.8) | 0.058 (9.8) | 0.293 (49.6) |
| Birbhum | 0.453 | 0.051** (11.3) | 0.223 (49.2) | 0.025*** (5.5) | 0.428 (94.5) | 0.153*** (33.8) | 0.22 (48.6) | -0.019** (-4.2) | 0.31 (68.4) |
| Barddhaman | 0.651 | -0.013*** (-2.0) | 0.321 (49.3) | 0.195*** (30.0) | 0.391 (60.1) | 0.141*** (21.7) | 0.404 (62.1) | 0.049 (7.5) | 0.42 (64.5) |
| Nadia | 0.671 | 0.114 (17.0) | 0.324 (48.3) | 0.086 (12.8) | 0.451 (67.2) | 0.172*** (25.6) | 0.447 (66.6) | 0.243*** (36.2) | 0.305 (45.5) |
| North 24 Parganas | 0.679 | 0.113 (16.6) | 0.331 (48.7) | 0.208*** (30.6) | 0.315 (46.4) | 0.047*** (6.9) | 0.463 (68.2) | 0.054 (8.0) | 0.415 (61.1) |
| Hooghly | 0.57 | 0.015*** (2.6) | 0.288 (50.5) | 0.097 (17.0) | 0.323 (56.7) | 0.077 (13.5) | 0.331 (58.1) | 0.069 (12.1) | 0.379 (66.5) |
| Bankura | 0.522 | 0.092 (17.6) | 0.251 (48.1) | 0.031*** (5.9) | 0.423 (81.0) | 0.086 (16.5) | 0.228 (43.7) | -0.027** (-5.2) | 0.482 (92.3) |
| Purulia | 0.545 | 0.258*** (47.3) | 0.228 (41.8) | 0.026*** (4.8) | 0.298 (54.7) | 0.078 (14.3) | 0.288 (52.8) | 0*** (0.0) | 0.545 (100.0) |
| Howrah | 0.752 | 0.07*** (9.3) | 0.37 (49.2) | 0.317*** (42.2) | 0.353 (46.9) | 0.065** (8.6) | 0.679 (90.3) | 0.108 (14.4) | 0.603 (80.2) |
| Kolkata | 0.743 | 0.215*** (28.9) | 0.341 (45.9) | 0 (0.0) | 0.743 (100.0) | 0.05*** (6.7) | 0.638 (85.9) | 0.024*** (3.2) | 0.621 (83.6) |
| South 24 Parganas | 0.618 | 0.048*** (7.8) | 0.314 (50.8) | 0.142*** (23.0) | 0.387 (62.6) | 0.084 (13.6) | 0.342 (55.3) | -0.002*** (-0.3) | 0.345 (55.8) |
| Paschim Medinipur | 0.711 | 0.096 (13.5) | 0.347 (48.8) | 0.093 (13.1) | 0.603 (84.8) | 0.173*** (24.3) | 0.483 (67.9) | 0.269*** (37.8) | 0.349 (49.1) |
| Purba Medinipur | 0.64 | 0.064** (10.0) | 0.313 (48.9) | 0.017*** (2.7) | 0.54 (84.4) | 0.071 (11.1) | 0.445 (69.5) | 0.054 (8.4) | 0.563 (88.0) |
| **West Bengal** | **0.67** | **0.053 (7.9)** | **0.334 (49.9)** | **0.104 (15.5)** | **0.38 (56.7)** | **0.059 (8.8)** | **0.431 (64.3)** | **0.038 (5.7)** | **0.406 (60.6)** |

Source: Author's calculation

Note: Figures in the parentheses are the corresponding percentage to total inequality of respective districts/State

| Sl No | Districts | Package component | Doctor's/surgeon's fee | Medicines | Diagnostic tests | Bed charges | Other medical expenses | Transport for patient | Other non-medical expenses |
|---|---|---|---|---|---|---|---|---|---|
| | Table 3: Disaggregation(%) of OOPE by different expenditure components for in-patient(hospitalisation) cases | | | | | | | | |
| 1 | Darjeeling | 57.0 | 6.0 | 12.6 | 6.6 | 4.7 | 4.1 | 3.7 | 5.4 |
| 2 | Jalpaiguri | 55.0 | 5.3 | 11.1 | 5.3 | 3.3 | 6.3 | 4.9 | 8.9 |
| 3 | Coochbehar | 14.4 | 20.0 | 24.3 | 9.9 | 4.8 | 7.6 | 8.4 | 10.6 |
| 4 | Uttar Dinajpur | 12.9 | 7.1 | 40.8 | 11.7 | 3.7 | 5.9 | 8.1 | 9.8 |
| 5 | Dakshin Dinajpur | 5.5 | 12.2 | 36.4 | 12.7 | 5.8 | 7.3 | 9.1 | 11.0 |
| 6 | Maldah | 21.6 | 9.8 | 23.8 | 12.4 | 6.5 | 5.4 | 10.6 | 9.9 |
| 7 | Murshidaba | 13.4 | 30.9 | 18.5 | 6.0 | 2.0 | 8.1 | 5.3 | 15.7 |
| 8 | Birbhum | 32.5 | 4.1 | 28.1 | 8.0 | 1.8 | 9.5 | 6.1 | 9.9 |
| 9 | Barddhaman | 29.9 | 7.1 | 28.5 | 11.8 | 4.1 | 6.9 | 5.1 | 6.6 |
| 10 | Nadia | 42.6 | 7.2 | 20.2 | 6.9 | 4.3 | 5.6 | 7.0 | 6.2 |
| 11 | North 24 Parganas | 44.7 | 8.2 | 18.2 | 8.7 | 5.5 | 5.8 | 4.9 | 3.9 |
| 12 | Hooghly | 57.3 | 3.5 | 15.1 | 5.6 | 2.7 | 6.4 | 4.5 | 5.0 |
| 13 | Bankura | 33.6 | 6.4 | 21.8 | 11.3 | 2.4 | 2.0 | 10.1 | 12.4 |
| 14 | Purulia | 19.3 | 11.6 | 18.7 | 10.9 | 8.4 | 7.5 | 11.2 | 12.4 |
| 15 | Howrah | 52.6 | 5.7 | 18.1 | 7.1 | 2.4 | 4.8 | 4.6 | 4.7 |
| 16 | Kolkata | 54.0 | 7.4 | 13.1 | 5.7 | 7.0 | 9.6 | 1.2 | 2.0 |
| 17 | South 24 Parganas | 30.7 | 13.1 | 23.9 | 14.1 | 5.1 | 5.1 | 3.5 | 4.5 |
| 18 | Paschim Medinipur | 41.6 | 8.9 | 19.7 | 10.6 | 5.2 | 3.6 | 4.3 | 6.1 |
| 19 | Purba Medinipur | 23.7 | 13.5 | 21.9 | 11.8 | 8.0 | 5.3 | 6.9 | 8.9 |
| | **West Bengal** | **38.2** | **38.2** | **10.2** | **20.1** | **9.0** | **4.7** | **6.4** | **4.8** |

Source: Author's calculation

| Sl. No | District | Doctor's/ surgeon's fee | Medicines_ AYUSH | Medicines Other than AYUSH | Diagonistic tests | Other medical expenses | Transport for patient | Other non-medical expenses |
|---|---|---|---|---|---|---|---|---|
| | **Table 4: Disaggregation(%) of OOPE by different expenditure components for Outpatient(OPD) cases** | | | | | | | |
| 1 | Darjiling | 10.60 | 0.48 | 76.53 | 4.72 | 1.22 | 2.53 | 3.92 |
| 2 | Jalpaiguri | 6.52 | 1.39 | 34.99 | 38.16 | 0.45 | 9.63 | 8.87 |
| 3 | Koch Bihar | 10.67 | 0.24 | 51.67 | 20.31 | 2.04 | 10.69 | 4.38 |
| 4 | Uttar Dinajpur | 12.04 | 2.34 | 72.82 | 5.23 | 1.05 | 4.73 | 1.79 |
| 5 | DakshinDinajpur | 16.60 | 1.99 | 56.87 | 15.87 | 0.46 | 6.53 | 1.68 |
| 6 | Maldah | 13.37 | 2.06 | 62.77 | 7.81 | 0.17 | 9.50 | 4.32 |
| 7 | Murshidabad | 13.30 | 4.63 | 57.18 | 6.78 | 0.57 | 6.85 | 10.70 |
| 8 | Birbhum | 17.32 | 0.28 | 58.93 | 2.74 | 2.27 | 10.08 | 8.37 |
| 9 | Barddhaman | 11.73 | 3.57 | 56.08 | 12.84 | 3.26 | 7.36 | 5.15 |
| 10 | Nadia | 19.48 | 0.92 | 63.99 | 4.55 | 0.27 | 6.52 | 4.26 |
| 11 | North 24 Parganas | 11.69 | 3.72 | 62.40 | 11.59 | 1.63 | 7.46 | 1.52 |
| 12 | Hugli | 13.67 | 2.90 | 65.97 | 6.58 | 4.15 | 4.90 | 1.83 |
| 13 | Bankura | 14.43 | 2.48 | 54.27 | 10.46 | 1.20 | 6.52 | 10.63 |
| 14 | Puruliya | 20.12 | 7.02 | 29.82 | 12.51 | 9.08 | 11.94 | 9.51 |
| 15 | Haora | 20.49 | 2.48 | 56.38 | 5.54 | 1.76 | 10.71 | 2.64 |
| 16 | Kolkata | 17.48 | 0.61 | 71.73 | 4.91 | 2.44 | 2.28 | 0.56 |
| 17 | South 24 Parganas | 12.51 | 2.40 | 67.02 | 6.84 | 1.78 | 6.12 | 3.32 |
| 18 | Paschim Medinipur | 8.98 | 1.04 | 67.13 | 9.16 | 1.89 | 7.37 | 4.42 |
| 19 | PurbaMedinipur | 18.03 | 1.21 | 59.15 | 8.86 | 0.89 | 5.88 | 5.97 |
| | **West Bengal** | **13.57** | **2.24** | **61.56** | **9.66** | **1.78** | **6.88** | **4.31** |

Source: Author's calculation